\documentclass[aps,prl,onecolumn,groupedaddress]{revtex4}
\usepackage{graphicx}
\usepackage{epstopdf}
\usepackage{hyperref}
\usepackage{amsmath,amsthm,amssymb,euscript,mathrsfs,epsf,epsfig}
\begin{document}

\title{ The Frenkel Line: a direct experimental evidence for the new thermodynamic boundary}

\author{Dima Bolmatov$^{1}$*}
\author{Mikhail Zhernenkov$^{1}$**}
\author{Dmitry Zav'yalov$^{2}$}
\author{Sergey N. Tkachev$^{3}$}
\author{Alessandro Cunsolo$^{1}$}
\author{Yong Q. Cai$^{1}$}

\affiliation{$^1$ National Synchrotron Light Source II, Brookhaven National Laboratory, Upton, NY 11973, USA}
\affiliation{$^2$ Volgograd State Technical University, Volgograd, 400005 Russia}
\affiliation{$^3$ Center for Advanced Radiation Sources, University of Chicago, Chicago, IL 60637, USA}

\begin{abstract}
Supercritical fluids play a significant role in elucidating fundamental aspects of liquid matter under extreme conditions. They have been extensively studied at pressures and temperatures relevant to various industrial applications. However, much less is known about the structural behaviour of supercritical fluids and no structural crossovers have been observed in static compression experiments in any temperature and pressure ranges beyond the critical point. The structure of supercritical state is currently perceived to be uniform everywhere on the pressure-temperature phase diagram, and to change only in a monotonic way even moving around the critical point, not only along isotherms or isobars. Conversely, we observe structural crossovers for the first time  in a deeply supercritical sample through diffraction measurements in a diamond anvil cell and discover a new thermodynamic boundary on the pressure-temperature diagram. We explain the existence of these crossovers in the framework of the phonon theory of liquids using molecular dynamics simulations. The obtained results are of prime importance since it implies a global reconsideration of the mere essence of the supercritical phase. Furthermore, this discovery may pave the way to new unexpected applications and to the exploration of exotic behaviour of confined fluids relevant to geo- and planetary sciences. \\
\\
* d.bolmatov@gmail.com, ** zherne@bnl.gov
\end{abstract}

\maketitle
Statistical mechanics is a very prominent part of physics \cite{llandau}. In the annals of statistical physics of aggregation states, last century and recent decades mark a very vibrant epoch \cite{aeinstein,pdebye,vwaals,mborn,jfrenkel,bwidom,nashcroft}. A series of successful macroscopic approaches suggests that a relatively simple microscopic theory should be capable of yielding realistic phase diagrams. Within each phase, the system is uniform in chemical composition and physical state. Critical point occurs under conditions of specific values of temperature, pressure and composition, where no phase boundaries exist \cite{bwidom1}. As the substance approaches critical temperature, the properties of its gas and liquid phases converge, resulting in only one phase at and beyond the critical point  -- a homogeneous supercritical fluid \cite{ekiren,barat}. Recently, it has been experimentally established that the supercritical fluids are, in fact, dynamically non-homogeneous in the neighborhood of the critical point \cite{tscopigno}. 

The understanding of the supercritical state has recently been revised \cite{annals2015}. It has been suggested to divide the supercritical state beyond the critical point and its neighborhood  into two distinct domains by introducing the Frenkel line in the framework of the unified phonon-based approach \cite{annals2015}. It has been shown that moving from one domain to another in the supercritical state is accompanied by changes in particle dynamics \cite{jpcl2015}. The Frenkel line separates these two regions beyond the critical point based on changes in phonon excitations, which defines the dynamic crossover \cite{jpcl2015}. Recently, {\it structural} \cite{dbolmatovjpc,dbolmatovjpcl} and {\it thermodynamic crossovers} \cite{dbolmatovnc} associated with the crossing of the Frenkel line have been theoretically predicted.  Importantly, the predicted structural and thermodynamic crossovers are closely related to fundamental changes in phonon states \cite{bolsr, dbolmatovsr}, providing new unexpected connections between elementary collective excitations and the structure \cite{dbolmatovjpc,dbolmatovjpcl}, thermodynamics and scaling laws of supercritical state \cite{dbolmatovnc}. 

\begin{figure}
	\centering
\includegraphics[scale=0.09]{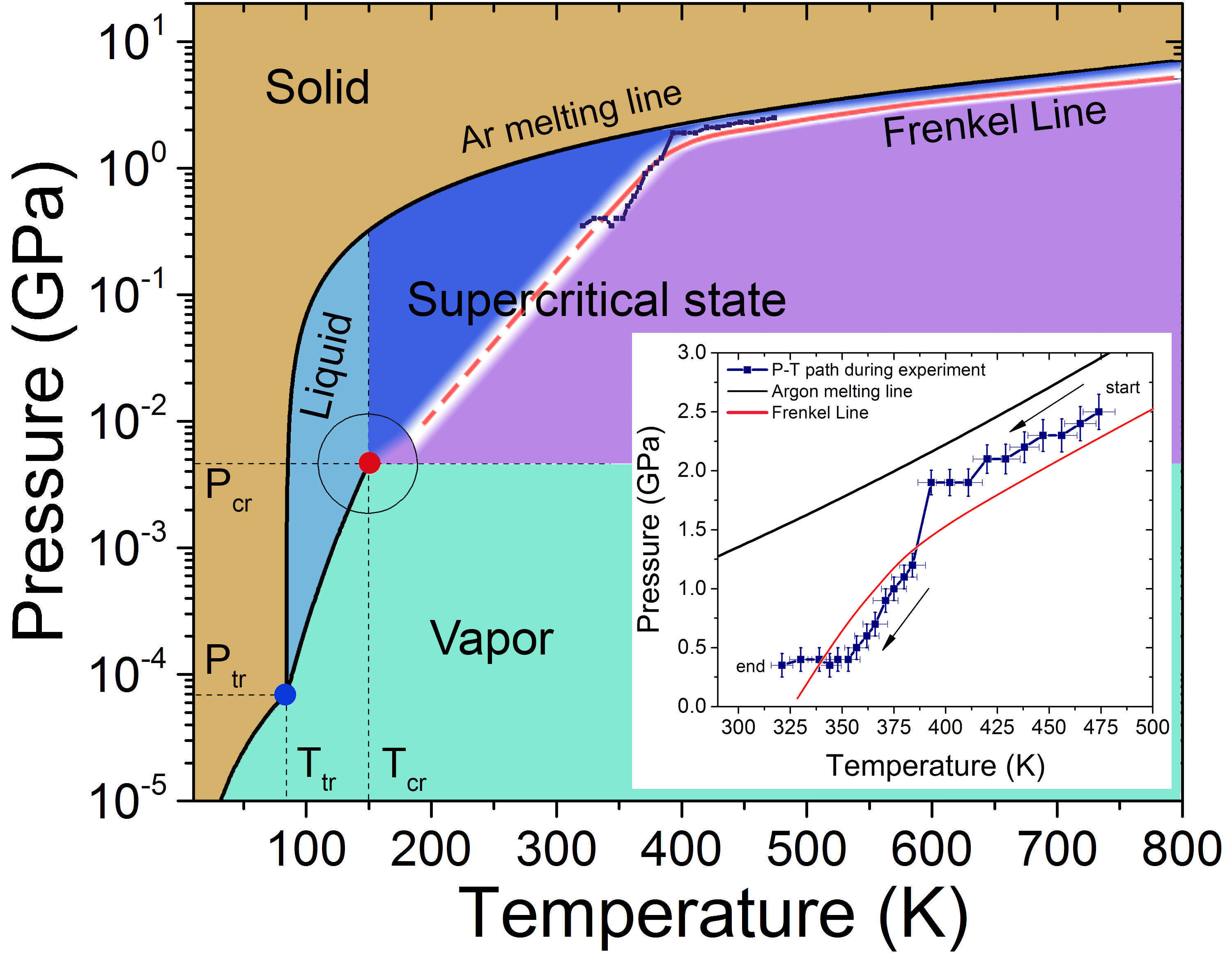}
\caption{{\bf Argon pressure-temperature phase diagram with the new thermodynamic boundary}. The phase diagram shows all major states of matter: solid, compressed liquid, supercritical and vapor phases. The inset displays the Argon melting line \cite{datchi}, the pressure-temperature {\it tour de experimental path} and conditions at which the Frenkel line was observed. The behavior of the Frenkel line below 0.4 GPa and 325 K is denoted by the dashed line and is to guide the eye only. The position of the Frenkel Iine at the vicinity (shown by the circle) of the critical point cannot be reliably determined as the critical point  strongly affects the behavior of all major thermodynamic quantities.  At higher P-T conditions, the Frenkel line is nearly parallel to the Ar melting line \cite{brazhka2013}.}
\label{diagram}
\end{figure}

In this work, we report results from a diffraction experiment on supercritical argon in a diamond anvil cell (DAC) which enables the observation of  structural transformations upon crossing the Frenkel line. These transitions correspond to a new thermodynamic boundary on argon pressure-temperature diagram (see Fig.~\ref{diagram}).
We explain the origin of the structural crossovers across the Frenkel line in the framework of the phonon theory of liquids using molecular dynamics (MD) simulations, and relate its origin to the change of thermodynamic properties of the supercritical state. The experimental results presented here are of great interest to the ongoing effort in elucidating various properties of disordered matter \cite{acunsolo, jeggert,lbenedict,kondorin,yheo,angel,larini,ktrachenko,adrozd,gorelli,dyre,bolstat,biroli,mauro,giordano,flen,tanaka,
vlevashov,adam,may}.
\section{Results}
\subsection{Mind the transverse phononic gaps}
Here, we introduce the Hamiltonian $H=H_{0}+H_{int}$, where $H_{0}$ defines a free theory with no interactions between phonons
\begin{equation}
\label{H_0}
H_{0}=\frac{1}{2}\sum_{\omega_q<\omega_{\rm D}}\left[\Pi_q^\alpha\Pi_{-q}^\alpha+\mu\omega_q^2 Q_q^\alpha Q_{-q}^\alpha\right]
\end{equation}
and $H_{int}$ is the interaction term that leads to spontaneous symmetry breaking
\begin{equation}
H_{int}=\sum_{\omega_q<\omega_{\rm D}}\left[-\frac{g}{2} |Q_{q}^{\alpha}|^4+\frac{\lambda}{6}|Q_{q}^{\alpha}|^6\right]
\end{equation}
$q$ is a multi-index $\{q_1,q_2,q_3\}$, $\omega_{\rm D}$ is the Debye frequency, and the parameter $\mu$ takes values 1 or 0. The parameters $g,\lambda\in\mathbb{R}^+$ are real non-negative couplings, as introduced in the theory of aggregation states \cite{bolsr}. $\Pi^\alpha_q$ and $Q_q^\alpha$ are the collective canonical coordinates and $|Q_q^\alpha|=(Q_{q}^\alpha Q_{-q}^\alpha)^{1/2}$. The configurations $\bar{Q}^\alpha_q$ and $\bar{\Pi}_q^\alpha$ minimise the energy of the system and break the $SO(3)$ symmetry to $SO(2)$. Minima of the potential
\begin{equation}
V[Q_q^\alpha]=\sum_{\omega_q<\omega_{\rm D}}\left[\frac{\mu}{2}\omega_q^2 |Q_q^\alpha|^2-\frac{g}{2} |Q_{q}^\alpha|^{4}+\frac{\lambda}{6}|Q_{q}^\alpha|^6\right]
\end{equation}
are found to be
\begin{equation}
\label{vac}
\begin{aligned}
|{Q}_q^\alpha|_{\pm}& =\left(\frac{g}{\lambda}+ \sqrt{\frac{\omega_{\rm F}^2-\omega_q^2}{\lambda}}\right)^{1/2},\\
|{Q}_q^\alpha|_{0}&=0,\\
\omega_{\rm F}& = \sqrt{\frac{g^2}{\lambda}}.
\end{aligned}
\end{equation}
$\omega_{\rm F}$ is the Frenkel frequency and defines the lower bound of the oscillation frequency of the atoms or molecules. It can be derived from the viscosity $\eta$ and shear modulus $G_{\infty}$ of a liquid \cite{dbolmatovsr}. Excitations of the phonon field around the ground state $\bar{Q}_q^\alpha$ can be written as
\begin{equation}
Q^\alpha_q=\bar{Q}_q^\alpha+\varphi_q^\alpha, \\  \alpha=1,2,3
\label{vacuum}
\end{equation}
where $\varphi_q^{2,3}$ and $\varphi^1_q$ are the transverse and the longitudinal modes respectively. For a chosen vacuum $\bar{Q}_q^\alpha=\delta^\alpha_1|\bar{Q}_q|$ we obtain the effective Hamiltonian
\begin{equation}
\label{Ham}
\begin{aligned}
H[\varphi_q] =\frac{1}{2}\sum_{0\leq\omega_q^{l,t,t}\leq\omega_{\rm D}}[\pi_q^{1}\pi_{-q}^{1}+\pi_q^{2}\pi_{-q}^{2}+\pi_q^{3}\pi_{-q}^{3}]+\\
\sum_{0\leq\omega_q^l\leq\omega_{\rm D}}\left[\frac{\omega_q^2}{2}\varphi^1_q\varphi^1_{-q}\right]+ \sum_{\omega_{\rm F}\leq\omega_q^{t,t}\leq\omega_{\rm D}}\left[\frac{\omega_q^2}{2}(\varphi^2_q\varphi^2_{-q}+\varphi^3_q\varphi^3_{-q})\right]
\end{aligned}
\end{equation}
where $l$ and $t$ stand for the longitudinal and transverse phonon polarizations, respectively. In the framework of the above formalism one may derive energy spectra in reciprocal space, varying the system parameters, which in real space can be attributed to different states of aggregation such as solids, liquids and gas \cite{bolsr}. In particular, this Hamiltonian predicts that heat capacity at constant volume per particle $c_V=\frac{1}{N}\left(\frac{\partial E}{\partial T}\right)_{\rm N}$ ($H\varphi=E\varphi$) drops down from approximately $3k_{\rm B}$ to about $2k_{\rm B}$  \cite{bolsr,dbolmatovsr}, where $k_{\rm B}$ is the Boltzmann constant. 

The effective Hamiltonian has a neat property regarding the low-frequency transverse phonon excitations in the {\it rigid}/compressed liquid regime (see the last term in Eq. \ref{Ham}), i.e. low-frequency wave-packets (long-wavelength limit) cannot be propagated in the {\it rigid liquids} due to the existence of the transverse phononic gaps in the spectrum ($0\neq\omega_{\rm F}<\omega_{q}^{t,t}<\omega_{\rm D}$). The inability to support low-frequency transverse  elementary collective excitations (low-energy cutoff) is a manifestation of the absence of the long-range order structural correlations in liquids (see DAC experiments) and {\it vice versa}. An increase in temperature leads to the disappearance of both the high-frequency transverse phonon modes ( $\omega_{\rm F}\xrightarrow{T}\omega_{\rm D}$, hence, $c_V$: $3k_{\rm B}\xrightarrow{T}2k_{\rm B}$) \cite{dbolmatovsr,dbolmatovprb,dbolmatovjap} and progressively the medium-range order pair correlations \cite{dbolmatovjpc, dbolmatovjpcl}. $c_V=2k_{\rm B}$ ( $\omega_{\rm F}=\omega_{\rm D}$, see Eq. \ref{Ham}) is the new thermodynamic limit( dubbed here the Frenkel line thermodynamic limit) along with other well-known $c_V=3k_{\rm B}$ (the Dulong-Petit law) and  $c_V=\frac{3}{2}k_{\rm B}$ (the ideal gas) thermodynamic limits which are also covered by Eq. \ref{Ham}. Therefore, crossing the Frenkel line results in fundamental changes of pair structural correlations both in reciprocal and real spaces \cite{dbolmatovjpc, dbolmatovjpcl}, as well as in the thermodynamics (heat capacity at constant volume $c_V$), scaling laws \cite{dbolmatovnc}, and phonon states in the supercritical matter \cite{bolsr,dbolmatovsr}.

Alternatively, the liquid energy can be calculated on the basis of pair correlations
\begin{equation}
\label{gas}
E=\frac{3}{2}Nk_{\rm B}T+2\pi N\rho\int_{0}^{\infty}r^{2}u(r)g(r)
\end{equation}
where $\rho$ is the density, $u(r)$ is the interatomic potential, and $g(r)$ is the pair distribution function which can also be represented as a  Fourier transform of the static structure factor $S(q)$ \cite{dbolmatovjpc}.

The first approach (see Eq. \ref{Ham}) takes into account the phonon contributions into the internal energy
of a liquid. As temperature increases ($\omega_{\rm F}\xrightarrow{T}\omega_{\rm D}$) the internal
energy and the heat capacity approach the Frenkel line thermodynamic limit meaning that $c_V=\left(\frac{1}{N}\frac{\partial E}{\partial T} \right)_{\rm N}\xrightarrow{T} 2k_{\rm B}$. On the other hand, from the second equivalent approach (see Eq. \ref{gas}), the internal energy with temperature increase loses the contribution from the medium range pair correlations which can be evidenced from disappearance of the
2$^{nd}$ $g(r)$ or $S(q)$ peaks. Thus, $c_V=\left(\frac{1}{N}\frac{\partial E}{\partial T} \right)_{\rm N}\xrightarrow{T} 2k_{\rm B}$. Therefore,  dynamic, structural and thermodynamic crossovers are interconnected  within the liquid and the supercritical state and governed by temperature
variations, hence, negligibly influenced by the pressure in terms of the described analysis.

It is noteworthy, that any approach what bears a "small parameter" and makes it possible to expand the energy of the system into the series, for example such as virial expansion, Meyer expansion, Percus-Yevick approach, hard-sphere model and other perturbation approaches are capable to describe weakly interacting systems, by the definition. As a result, these
approaches fail in describing the systems with strong interactions like compressed liquids. Hence, these methods are unable to predict and describe the structural crossover between compressed and non-compressed liquid regimes which is described above and the subject of this work.

\begin{figure*}[htp]
  \centering
  \begin{tabular}{cc}
    \includegraphics[scale=0.15]{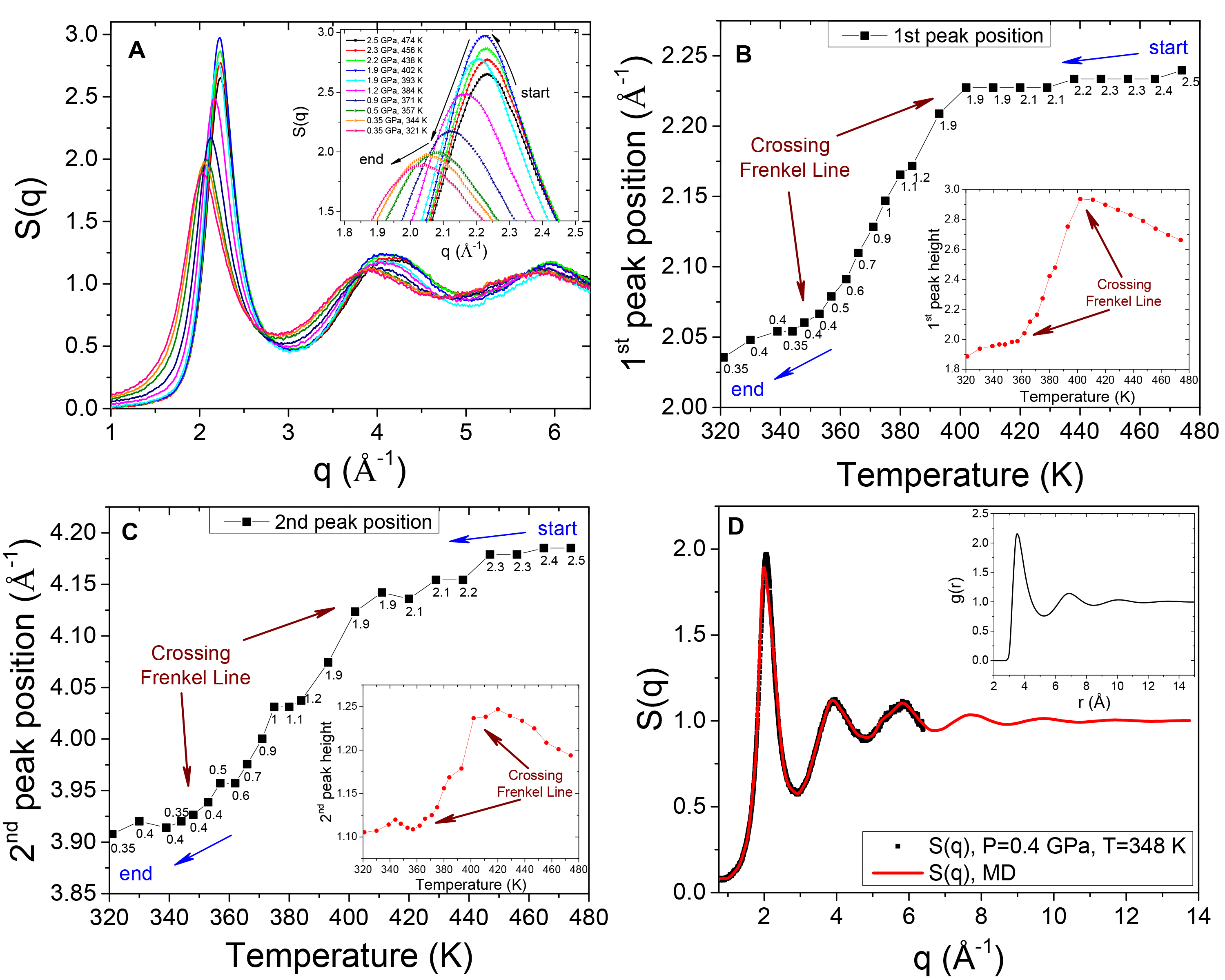}




  \end{tabular}
\caption{{\bf Experimental reciprocal space evolution of pair correlations across the Frenkel line}.({\bf A}) The measured static structure factor $S(q)$ on X-ray diffraction experiments at different pressure-temperature conditions. ({\bf B, C}) First and second  diffraction $S(q)$ peaks pressure-temperature position variation. Labels indicate the pressure in GPa units. Insets show peaks pressure-temperature heights variation illustrating non-uniform behaviour, hitherto unanticipated \cite{ekiren}. ({\bf D}) Comparison of the experimentally obtained $S(q)$ with the $S(q)$ calculated from MD simulations. The inset shows the $g(r)$ derived from the MD simulations.}
\label{figsoq}
\end{figure*}
\subsection{Diamond anvil cell experiments}
The combination of synchrotron X-ray facilities with high-pressure methods provides new experimental tools for probing the structure and dynamics of materials. In our experiment, we performed high pressure/high temperature X-ray diffraction (XRD) measurements using a BX90 (DAC) \cite{kantor} at GSECARS beam line 13-ID-D of the Advanced Photon Source (APS), Argonne National Laboratory. The DAC was used in combination with tungsten-carbide seats and full diamond anvils with a 500 $\mu$m culet size.  250 $\mu$m-thick rhenium gasket was  pre-indented to a thickness of about 40 $\mu$m. A hole with a diameter of about 120 $\mu$m was drilled in the middle of the pre-indented area. Conventional resistive heating was used to heat the sample up to 500 K. The temperature was measured by two thermocouples attached to the diamond surface at the vicinity of the culet. The $^{\rm 40}$Ar was loaded using a COMPRES/GSECARS gas-loading system at APS \cite{compress} up to initial pressure of 1 GPa. A ruby sphere was used for the pressure calibration \cite{sandeep}. Given the Ar loading pressure, the thickness of the gasket’s pre-indented area and the diameter of the drilled hole, the sample would amount to approximately 10$^{16}$ Ar atoms. The sample was measured at the photon energy of 37.077 keV; the XRD patterns were recorded by a MAR-165 CCD camera with 79x79 $\mu$m$^2$ pixel size with the exposure time of 10 seconds per XRD pattern. Following every temperature change, the DAC was allowed to equilibrate for, at least, 5 minutes before the XRD pattern was collected. The raw XRD patterns were further processed using the Dioptas software package. The background measured from the empty cell was subtracted from each spectrum. Each $S(q)$ curve was  normalized to the Ar compressibility limit (see, e.g. page 31, \cite{mcdonald}) for a given pressure and temperature. The $S(0)$ values are obtained from independent thermodynamic compressibility data (see the NIST database).

In the present definitions of supercritical state \cite{ekiren,barat}, even moving around the critical point \cite{ekiren,barat} on the pressure-temperature diagram (not only along isotherms or isobars) can be considered appropriate to detect possible structural non-uniformities based on pair correlations analysis \cite{ekiren,barat}. Selected $S(q)$ curves at different P-T conditions for wavenumbers $q$ up to 6.5 \AA$^{-1}$ are presented in Fig.~\ref{figsoq}a. In Fig. \ref{figsoq}(b,c), we present the position variation of first and second $S(q)$ peaks as a function of temperature. Insets show the $S(q)$ peak height as a function of temperature illustrating the non-uniform behaviour consistent with previous theoretically predicted results \cite{dbolmatovjpc,dbolmatovjpcl}.
\begin{figure}
	\centering
\includegraphics[scale=0.2]{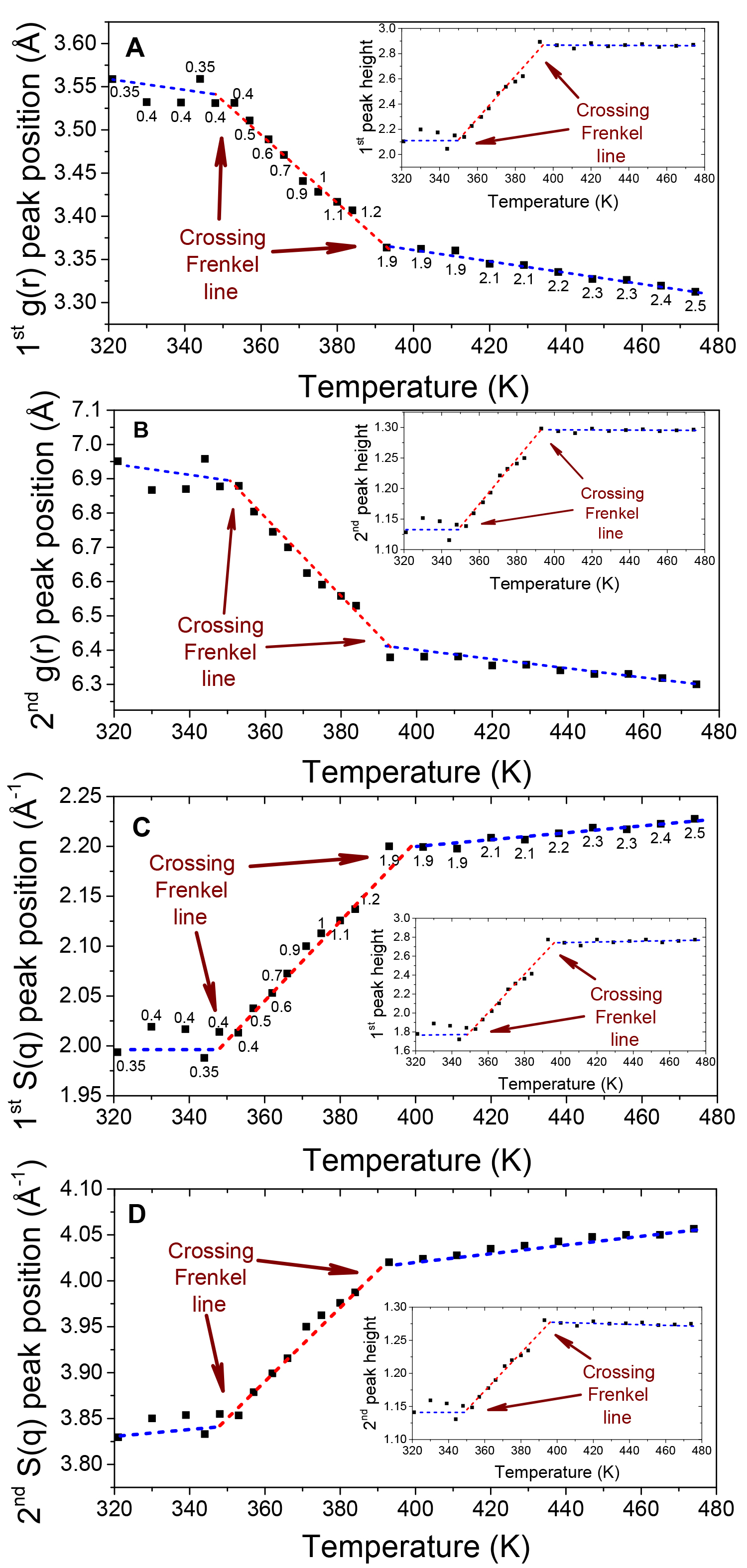}



\caption{{\bf Real space evolution of pair correlations  across the Frenkel line as derived from MD simulations}.({\bf A, B}) and ({\bf C, D}) first and second  $g(r)$ and $S(q)$ peaks pressure-temperature position variation derived from the MD simulations at experimental conditions, respectively. ({\bf A, C}) labels indicate the pressure in GPa units. Insets show $g(r)$ and $S(q)$ peaks pressure-temperature heights variation exhibiting heterogeneous behaviour, hitherto unanticipated \cite{ekiren}. The dashed lines are to guide the eye only.}
\label{figgor}
\end{figure}

We would like to stress that the evolution of the $S(q)$ peak positions (see Figs.~\ref{figsoq}(b,c)) as a function of just pressure is featurelessly monotonic and has no implications for the observation of the Frenkel line. As we stated above, the thermodynamics of supercritical fluids shall be studied as a function of temperature with a little regard to the pressure deviations along the experimental path. Indeed, as can be seen from the previous studies \cite{gorelik-pidaras} the dependence of the first and the second Ar $S(q)$ peak positions as a function of pressure at the fixed temperature (isothermal scan) is essentially $linear$.
Clearly,  the mere $linear$ variation of the $S(q)$ peak positions as function of pressure cannot affect the complex evolution of the peak positions presented in the Figs.~\ref{figsoq}(b,c)(also see Supplementary Materials). 

Therefore, the structural crossover can only be observed as a function of temperature (in contrast to previous works \cite{gorelik-pidaras} where isothermal scans were studied and no crossovers detected) and explained within the framework of the phonon theory of liquids where temperature (not the pressure) is the key variable.

\subsection{MD simulations at experimental conditions}
In order to analyze the pair correlations in real space, see Fig. \ref{figgor}(a,b), we performed MD simulations at experimental conditions. In addition, we compared the experimentally determined $S(q)$ with $S(q)$ derived from MD simulations showing good agreement, see Fig.~\ref{figsoq}d, and consistent temperature evolution, see Figs.~\ref{figsoq}(b,c)-~\ref{figgor}(c,d). The  structure factor $S(q)$ can be defined as
\begin{eqnarray}
S(q) = 1 + 4\pi\varrho \int_{{0}}^{{R_{max}}}dr r^2{\frac{{\sin{qr}}}{{qr}}} \left({g(r)-1}\right)
\end{eqnarray}
where the $g(r)$ is the pair distribution function \cite{nashcroft}, it describes the distribution of distances between pairs of particles contained within a given volume, and $R_{max}$  is the distance cutoff parameter set to 20 \AA \ which we have found to be sufficient to converge the integral. We have used LAMMPS simulation code to run a Lennard-Jones (LJ, $\epsilon$/k$_{\rm B}$=119.8 K, $\sigma=$3.405) fluid fitted to Ar properties \cite{skoenig} with 32678 atoms in the isothermal-isobaric (NPT) ensemble. We have used 3000 processors of a high-throughput cluster with a runtime over 300 picoseconds.

In Fig. \ref{figgor} we observe the alternation of regimes, which is a result of power laws change, both for the first and the second peak positions and heights resulting from distinct changes between medium- and short-range order correlations \cite{dbolmatovjpc,dbolmatovnc}.  Different regimes are manifested in different slopes of peak positions and their heights as a function of temperature (red and blue dashed lines are guide for the eyes only). The current interchange of the medium- and short-range order correlations is reflected in the last term in Eq. \ref{Ham}. The presence or the absence of the last term in the Hamiltonian  describes structure and thermodynamics above and below the Frenkel line, respectively, which can be referred to as different regimes. Remarkably, experimental evolution of pair correlations (see Fig.\ref{figsoq}) also illustrates non-uniform behavior as simulated does (see Fig.~\ref{figgor}). The non-uniform behaviour of pair correlations both in reciprocal  and in real  spaces is closely related to the differences in the relaxation processes, phonon excitations and thermodynamics above and below the Frenkel line \cite{bolsr,dbolmatovnc}. 
\begin{figure*}[htp]
  \centering
  \begin{tabular}{cc}
  
   \includegraphics[scale=0.15]{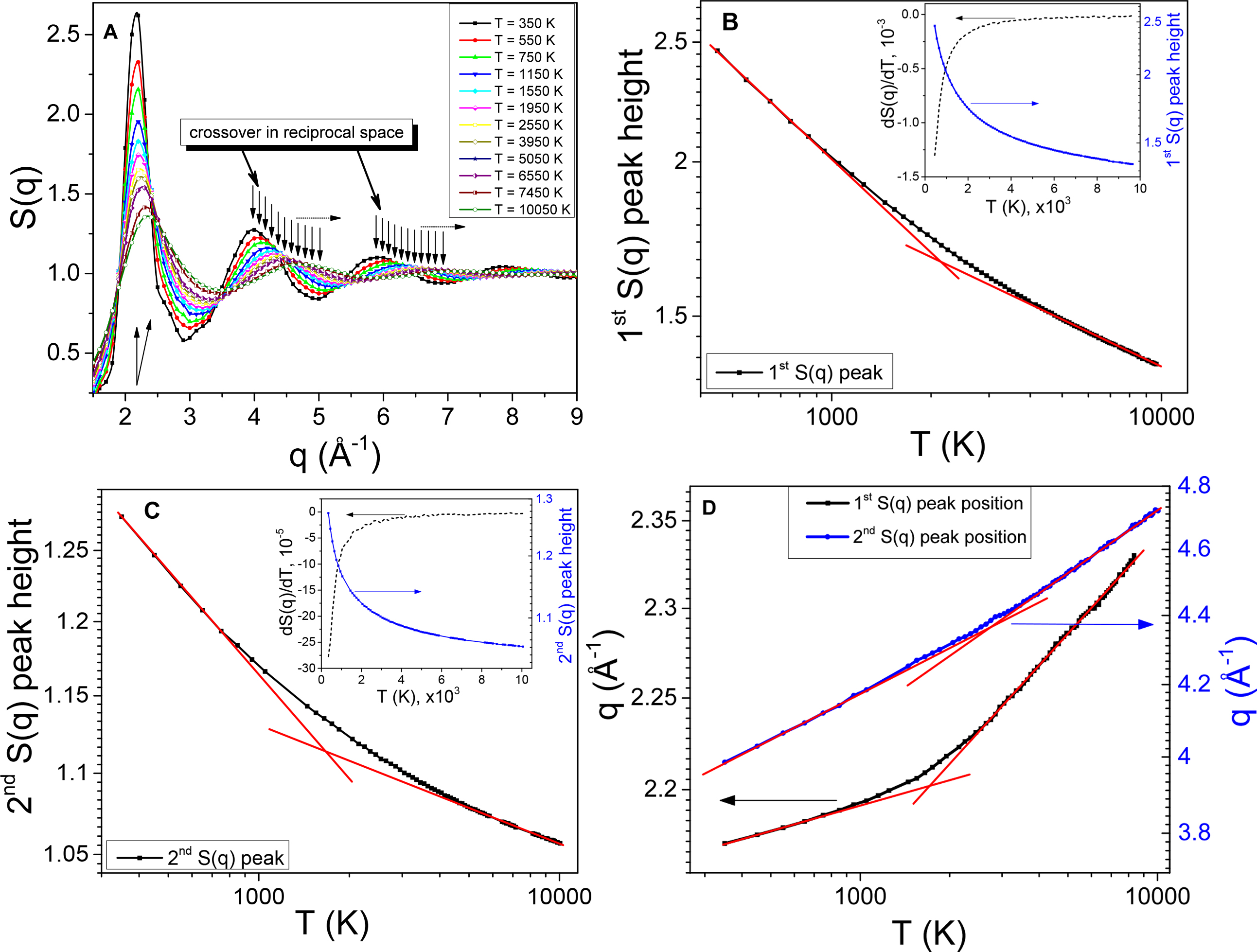}

  \end{tabular}
 \caption{{\bf MD simulations: structure temperature variation in reciprocal space}. ({\bf A}) Evolution of static structure factor $S(q)$ showing the disappearance of the medium-range correlations with increasing temperature. ({\bf B,C }) First  and second $S(q)$ peaks heights temperature evolution. Insets show $S(q)$ peaks and their derivatives plotted in linear scale. Arrows in the insets indicate the corresponding Y-axis for each curve. ({\bf D}) The position of the 1$^{st}$ and 2$^{nd}$ $S(q)$ peaks as a function of temperature. Arrows indicate the corresponding Y-axis for each curve. ({\bf B, C, D}) The graphs highlight two different regimes with two unique power laws (denoted by red solid straight lines) governing $S(q)$ peaks positions and heights resulting from the crossing the Frenkel line.}
\label{figfo}
\end{figure*}

Although the structural order of a fluid is usually enhanced by isothermal compression or isochoric cooling,
a few notable systems show the opposite behaviours. Specifically, for example for H$_{2}$O \cite{bett}, increasing the system density can disrupt the structure of liquid-like fluids, while lowering temperature or strengthening attractive interactions can weaken the correlations of fluids with short-range attractions. Structural order is a particularly useful measure of the observed changes crossing the Frenkel line. It also correlates strongly with the static structure factor and the pair distribution function behaviour  which provides insight into the structural crossover in the supercritical state. At elevated temperatures ($>$ 1000 K), decrease of the first (both $S(q)$ and $g(r)$) peak and the near disappearance of the second and third peaks imply  the medium-range order correlations are no longer present. Such high temperatures are not readily accessible in DAC experiments but can easily be studied using MD simulations. Nevertheless, the experimental evidence for this behavior is clearly seen as a decrease of the 1$^{st}$ and 2$^{nd}$ $S(q)$ peak heights above the T $\sim$ 410 K as shown in Fig. \ref{figsoq}(b,c) (see insets). 
The peak height decrease is also presented in Fig. \ref{figfo} and Fig. \ref{figfi}. The temperature variation of the 1$^{st}$ and 2$^{nd}$ $S(q)$ peak positions is also similar in the DAC experiment and MD simulations in the temperature range above T $\sim$ 410 K. However, the structural crossover at elevated temperatures cannot be observed experimentally in DAC measurements as the transition takes place at $\sim$ 1500 K.
\subsection{MD simulations at extended temperature range}

\begin{figure*}[htp]
  \centering
  \begin{tabular}{cc}
    
 \includegraphics[width=125mm]{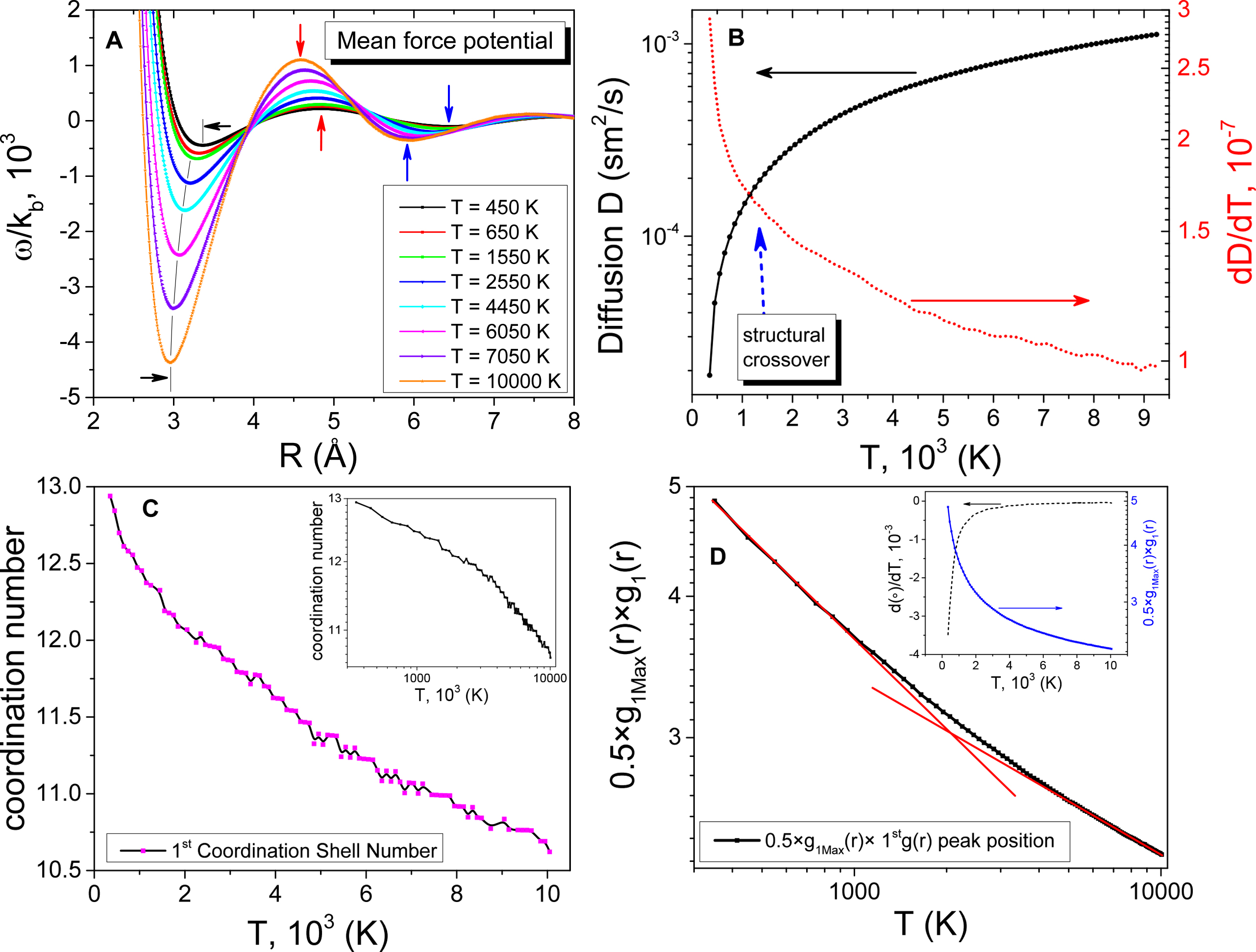}

  \end{tabular}
\caption{{\bf MD simulations: structure temperature variation in real space}. ({\bf A}) Evolution of mean force potential, ({\bf B}) diffusion and ({\bf C}) its first derivative  and first coordination shell  at different temperatures. ({\bf D}) $(\circ)=0.5\times g_{1Max}(r)\times$ $1^{st}$ peak position is plotted in Log-Log scale and highlights two different regimes with two unique power laws (represented by red solid straight lines). Inset shows the first coordination shell plot in Log-Log scale. Arrows indicate the corresponding Y-axis for each curve.}
 \label{figfi}
\end{figure*}

To analyze the behaviour of pair correlations in detail above and below the Frenkel line we run the MD simulations within very wide temperature range, which is not accessible for DAC experiments. Here, we have simulated a one-component Lennard-Jones (LJ, $\epsilon=$0.994, $\sigma=$3.405) fluid fitted to Ar properties \cite{martin}. The system is composed of a constant-volume (NVE) ensemble of 32000 atoms. Simulations were performed over thewide temperature range extending well into the supercritical region (see Fig.~\ref{figfo}). The temperature range used is located approximately between 3$T_c$ and 167$T_c$, where $T_c$  is the critical temperature of Ar, $T_{c}\simeq$150 K or 1.3 in LJ units. The system was equilibrated at constant temperature. The simulated density, 1880 kg/m$^3$ (1.05 in LJ units), corresponds to approximately three times the critical density of Ar. A typical MD simulation lasted about 50 picoseconds (ps), where the pair distribution function $g(r)$, static structure factor $S(q)$ and self-diffusion were averaged over the last 20 ps of simulation, after 30 ps of equilibration. A characteristic relaxation time of any liquid, including Ar, is about 0.1-0.3 ps, therefore the MD simulations significantly exceeded the mean collision time. Simulations were performed at 100 temperature points within the temperature range of interest. The temperature variation of these quantities is shown in Figs.~\ref{figfo}-\ref{figfi}. 

In the limiting case of no interaction, the system behaves as an ideal gas and the structure factor is completely featureless: $S(q)=1$. There is no correlation between the positions $\mathbf{r}_j$ and $\mathbf{r}_k$ of different particles (they are independent random variables) so the sum over the off-diagonal terms in equation
\begin{eqnarray}
S(q) = 1 + \frac{1}{N} \left \langle \sum_{j \neq k} \mathrm{e}^{-i \mathbf{q} (\mathbf{r}_j - \mathbf{r}_k)} \right \rangle
\end{eqnarray}
vanishes, i.e.
\begin{eqnarray}
\langle \exp [-i \mathbf{q} (\mathbf{r}_j - \mathbf{r}_k)]\rangle = \langle \exp (-i \mathbf{q} \mathbf{r}_j) \rangle \langle \exp (i \mathbf{q} \mathbf{r}_k) \rangle = 0
\end{eqnarray}
Even for interacting particles at high magnitudes of the scattering vector $\bf{q}$, the structure factor goes to 1. This result follows from equation 

\begin{eqnarray}
 S(q) = 1 + \rho \int_V \mathrm{d} \mathbf{r} \, \mathrm{e}^{-i \mathbf{q}\mathbf{r}} g(r)
\end{eqnarray}
since $S(q)-1$ is the Fourier transform of the $g(r)$ and thus goes to zero for high values of the argument $q$. This reasoning does not hold for a perfect crystal, where the distribution function exhibits infinitely  peaks. To examine the temperature changes of $S(q)$ in more detail (see Fig.~\ref{figfo}a), we show temperature variation of the heights and positions of the first and second peaks of $S(q)$ in Fig.~\ref{figfo}(b--d). There is a steep decrease of both peaks at high temperature, Fig.~\ref{figfo}(b,c), followed by their flattening at high temperature, with the crossover between the two regimes taking place around 1500 K. The crossover is visible in the insets of Fig.~\ref{figfo}(b,c) where we plot the temperature derivative of the heights of both peaks. These plots clearly show two structural regimes corresponding to the fast and slow changes of the $S(q)$ peaks and their positions. The structural crossover in the reciprocal space of supercritical state at elevated temperatures which we observe in Figs.~\ref{figfo}(b--d), supports the experimental results from the main text. It should be noted, that the observed crossover is not related to the melting line, as the $S(q)$ profile of the Ar solid phase (well-ordered phase) would exhibit very clear sharp peaks throughout the $q$-range covered by MD simulations. In contrast, the 1$^{st}$ and the 2$^{nd}$ S(q) peaks become wider and weaken with temperature increase signifying the strong disorder.

The structural crossover is further evidenced by the calculation of mean force potential (MFP), self-diffusion and first coordination shell parameters that include the first shell coordination number, the height and the first $g(r)$ peak position at different temperatures (see Fig.~\ref{figfi}). The temperature variations of the system properties are manifested in both real (Fig.~\ref{figfi}) and reciprocal (Fig.~\ref{figfo}(b--d)) spaces.

Various methods have been proposed for calculating the MFP. The simplest representation of the MFP is the separation $r$ between two particles as the reaction coordinate. The MFP (see Fig.~\ref{figfi}a) is related to the $g(r)$ using the following expression for the Helmholtz free energy \cite{kirkwood}
\begin{eqnarray}
 w=-k_{\rm B}T\ln{\left[g(r)\right]}
\end{eqnarray}
where $k_{\rm B}$ is the Boltzmann constant. We use Einstein's formula to calculate self-diffusion from the mean square distance (MSD) travelled by a certain particle over a certain time interval. At the limit of infinite observation time, self-diffusion in terms of MSD \cite{heyes1} becomes
\begin{eqnarray}
D=\lim_{t\rightarrow\infty}\frac{1}{6Nt}\left<\sum_{i=1}^{N}[\vec{r}_{i}(t)-\vec{r}_{i}(0)]^2\right>
\end{eqnarray}
where $\vec{r}_{i}$ is displacement vector of the $i$-th atom at $t$ time and the term $[\vec{r}_{i}(t)-\vec{r}_{i}(0)]^2$ is the MSD.

\section{Discussion}
The relationship between system properties in the real and reciprocal spaces is of prime interest in condensed matter physics. It is widely recognised that such a relationship may exist in some classes of systems but not in others. In this work, the DAC experiments and MD simulations show that not only the supercritical state is physically non-uniform  but also that manifestations of the non-uniformity are observed in both real and reciprocal spaces ($S(q)$ and $g(r)$ peaks exhibit a "zigzag" behavior). We relate the structural crossovers in the supercritical state to changes in elementary collective excitations, power laws and thermodynamics in the crossing of the Frenkel line. Crossing the Frenkel line corresponds to a quantitative change of the supercritical fluid's atomic structure and the transition of the substance from the compressed liquid structure to the non-compressed gas-like structure (see Figures \ref{figsoq}-\ref{figgor} and discussion below). 

The structural crossover takes place continuously when the liquid relaxation time $\tau$ (the average time between two consecutive atomic jumps at one point in space \cite{bolsr,dbolmatovsr}) approaches its minimal value $\tau_{\rm D}$, the Debye vibrational period, upon which the system loses its ability to support high-frequency propagating shear modes with $\omega >\frac{2\pi}{\tau}$ and behaves like a gas \cite{dbolmatovjpc,dbolmatovjpcl,dbolmatovnc}. When all shear modes are lost, only the longitudinal mode remains in the system and the heat capacity at constant volume becomes $c_{V}=2 k_{\rm B}$ per particle \cite{dbolmatovjpc,dbolmatovnc}. This result can be easily obtained from the phonon theory of liquids in the classical limit \cite{bolsr}. Therefore, the observed thermodynamic boundary in the DAC experiments is also closely related to both dynamic and thermodynamic crossovers existing in the supercritical state \cite{dbolmatovjpc,dbolmatovnc}.

The thermodynamic boundary discovered in this work has several possible implications to astrobiology and  the existence of alien life with the impact of extreme conditions on biomolecules \cite{oullrich}. Astonishingly, a number of species of bacteria are tolerant of supercritical CO$_{2}$ and can survive under severe conditions of pressure and temperature \cite{nbudisa}. Recently, we have studied structural properties of the supercritical CO$_{2}$ \cite{dbolmatovjpcl} where we provided the evidence for the existence of persistent medium-range order correlations that make supercritical CO$_{2}$ nonuniform and heterogeneous on an intermediate length scale. In the first shell of the CO$_{2}$ cluster both carbon and oxygen atoms experience gas-like type correlations with short-range order interactions while within the second shell, oxygen atoms essentially exhibit a liquid-like type of correlations due to localization of transverse-like wave-packets. Atoms inside the nearest-neighbor heterogeneity shell play a catalytic role providing a mechanism for diffusion on an intermediate length scale. Extraterrestrial organisms might use these peculiar structural and thermodynamic advantageous properties of the supercritical CO$_{2}$ to survive biologically. Other possibilities include exoplanets with 2-5 times the mass of the Earth with stronger gravitational pulls (super-Earths) and, thus, having supercritical atmospheres and/or oceans \cite{dbolmatovjpcl}. Therefore, further theoretical and experimental study of thermodynamics, dynamics and structure of various supercritical fluids and supercritical carbon dioxide in particular are in a strong demand. We believe that this discovery will boost the  industrial use of supercritical fluids in more efficient way, assist us to search for other  alternative conditions for an extraterrestrial fine-tuned life and also can lead to greater understanding in another disordered systems such as glasses and granular materials.

\section{Acknowledgements} The work at the National Synchrotron Light Source-II, Brookhaven National Laboratory, was supported by the U.S. Department of Energy, Office of Science, Office of Basic Energy
Sciences, under Contract No. DE-SC00112704. Synchrotron experiment was performed at GeoSoilEnviroCARS (Sector 13), Advanced Photon Source (APS), Argonne National Laboratory. GeoSoilEnviroCARS is supported by the National Science Foundation - Earth Sciences (EAR-0622171), Department of Energy - Geosciences (DE-FG02-94ER14466) and the State of Illinois. Use of the Advanced Photon Source was supported by the U. S. Department of Energy, Office of Science, Office of Basic Energy Sciences, under Contract No. DE-AC02-06CH11357 We are grateful to Neil Ashcroft, Joel Lebowitz, Giovanni Jona-Lasinio, Jerome Percus, John Wheeler, Edvard Musaev, Oleg Kogan, Gilberto Fabbris, Ivar Martin, Colin Wilson and Zeb Kramer for inspiring discussions.

\section{Author contributions}
D.B., M.Z. and Y.Q.C. designed the research. M.Z., D.B. and Y. Q. C. prepared the high pressure cell and performed resistive heating experiment. S.N.T. performed Ar gas loading. D.B., D.Z., A.C., Y.Q.C. and M. Z. performed the numerical simulations.  D.B. and M.Z. wrote the manuscript. All authors discussed the results and commented on the manuscript.

\section{Competing financial interests}

The authors declare no competing financial interests.

\end{document}